\documentstyle[seceq,preprint]{jpsj}
\renewcommand{\theequation}{\arabic{equation}}

\def\runtitle{Optical Conductivity and Metal-Insulator Transition 
of the 2D Hubbard Model}
\def\runauthor{Hiroki {\sc Nakano} and Masatoshi {\sc Imada}}

\title
{Optical Conductivity 
of the Two-Dimensional Hubbard Model
}

\author
{ 
Hiroki {\sc Nakano}
and Masatoshi {\sc Imada}
}

\inst
{
Institute for Solid State Physics, University of Tokyo, 
Roppongi 7-22-1, Minato-ku, Tokyo 106-8666\\
}

\recdate
{
}

\abst
{
Charge dynamics of the two-dimensional Hubbard model is investigated.
Lancz$\ddot{\rm o}$s-diagonalization results 
for the optical conductivity and the Drude weight 
of this model are presented.  
Near the Mott transition, large incoherence
below the upper-Hubbard band is obtained
together with a remarkably suppressed Drude weight in two dimensions
while the clearly coherent character is shown in one dimension.  
The two-dimensional results are consistent with previous results
from quantum Monte Carlo calculations indicating that
the Mott transition in this two-dimensional model
belongs to the universality class characterized
by the dynamical exponent of $z=4$.
}

\kword
{
two-dimensional Hubbard model, 
Mott transition, optical conductivity, Drude weight, 
incoherence, exact diagonalization, dynamical exponent
}

\begin{document}
\sloppy
\hyphenation{Hamil-ton-ian}
\maketitle

Physics of metal-insulator transitions is one of the major issues in 
current condensed-matter physics 
from both experimental and  theoretical viewpoints\cite{Imada_review}. 
The Hubbard model is one of  the simplest models 
which can describe the metal-insulator transition. 
This model shows a Mott transition originating in strong 
correlation between electrons.  
Unfortunately, 
properties of this model in two and three dimensions 
are not completely understood. 

 According to the scaling hypothesis  
for the Mott transition\cite{Imadad_scaling}, 
the Drude weight and the compressibility have 
the critical dependence on the doping concentration $\delta$ as 
$D \propto \delta^{1+(z-2)/d}$ and $\kappa \propto \delta^{1-z/d}$, 
respectively, where $z$ denotes the dynamical exponent and 
$d$ represents the spatial dimension. 
The transition is characterized by $z$.  
The value of $z$ may depend on systems and the scaling theory (ST) itself 
does not determine it. 
To fix the value of $z$, one has to estimate it explicitly, for example, 
using numerical results. 
Dagotto {\it et al.} reported $D \propto \delta$ 
for the two-dimensional (2D) Hubbard model\cite{Dagotto_Hub}. 
If one employs the ST, 
$D \propto \delta$ suggests $z=2$, 
which is in the same universality class as 
the band metal-insulator transition. 
From the quantum Monte Carlo (QMC) calculations\cite{Furukawa_Imada}, 
on the other hand, the chemical potential $\mu$ is scaled by $\delta^2$. 
This leads to $\kappa \propto 1/\delta$ and $z=4$. 
Another QMC calculation for the chemical-potential dependence 
of the localization length, $\xi_l$, shows 
$\xi_l$=$|\mu - \mu_{c}|^{-\nu}$ 
with $\nu=\frac{1}{4}$\cite{Assaad_Imada},   
which also suggests 
$z=4$. 
Thus, if one is based on the ST, 
a discrepancy remains between these two QMC results and 
the above result by Dagotto {\it et al.}, which could 
pose a suspicion on the validity of the ST. 

Quite recently, it was reported that $D$ 
in the 2D $t$-$J$ model has a very small value near the Mott transition, 
which strongly indicates $z=4$\cite{Tsune_Imada}. 
The critical behavior of $\kappa$\mbox{\,}$\propto$\mbox{\,}$1/\delta$ 
in this model was also reported\cite{Kohno}. 
Since the $t$-$J$ model is obtained in the strong-coupling limit of 
the Hubbard one in small {\it J}/{\it t}, 
it is widely believed that these models 
show the same critical behavior at least at realistic values 
of {\it J}/{\it t}.  
To have further insight on this issue, in our present analyses , 
we treat the finite-size (FS) effect more carefully than 
in the literature and reexamine 
the optical conductivity 
of the 2D Hubbard model. 

We study a model given by 
$
{\cal H}$=$-t \sum_{\langle i j \rangle , \sigma}
c^{\dagger}_{i\sigma} 
c_{j\sigma} $
 $+ U\!\sum_{i}\!n_{i{\tiny \uparrow}} n_{i\downarrow}, 
$
where the creation(annihilation) of an elec\-tron at site $i$ 
with spin $\sigma$ is denoted by $c^{\dagger}_{i\sigma}$($c_{i\sigma}$) 
with the number operator $n_{i\sigma}$. 
Here, we treat only the nearest-neighbor hopping 
with amplitude {\it t}\mbox{\,}=1. 
The doping concentration is given by 
$\delta$=1$-\frac{N_{\rm e}}{N_{\rm s}}$, where 
$N_{\rm e}$ and $N_{\rm s}$ denote the number of electrons and 
sites, respectively. 
We choose a boundary condition (BC)
among periodic, anti-periodic and mixed\cite{mixed} 
ones so that the ground-state energy becomes 
as low as possible as it should be. 
In fact, with this procedure, 
negative and unphysical values of $D$ 
obtained in Ref.\mbox{\,}\ref{Dagotto_Hub} 
near half filling do not appear 
because $D$ is the curvature of the flux dependent energy 
at the true ground state. 
The FS effect in this procedure is expected to be smaller 
than the fixed BC. 
Actually, when one considers the sum of squared deviations 
from the thermodynamic limit in $D$ 
 of 4$\times$4-site system with even $N_{\rm e}$ at $U$=0, 
the sum for the optimized BC is $\sim$0.0021, smaller than 
$\sim$0.0032 for the BC fixed to be periodic. 

  We use exact diagonalization of an FS cluster 
of $4 \times 4$ sites  
based on Lancz$\ddot{\rm o}$s algorithm and 
continued-fraction-expansion method\cite{frac_expansion}. 
Hilbert space of this cluster has a huge dimension 
of order of $10^8$. 
To treat the dimension, we carry out 
the parallel processing on supercomputers 
with a distributed-memory system. 

  We here calculate the optical conductivity defined as 
$\sigma (\omega) =[\sigma_x (\omega)+\sigma_y (\omega)]/2$ 
with 
$\sigma_{\alpha} (\omega) 
$ given by
$2\pi e^2 D_{\alpha} \delta(\omega) $ 
+
$\frac{\pi e^2}{N} 
\sum_{n(\ne 0)}\frac{|\langle n|j_{\alpha}|0\rangle |^{2}}{E_n-E_0}
\delta(\omega -E_n+E_0)$. 
Here $D_{\alpha}$ is the Drude weight 
along $\alpha$-direction ($\alpha$={\it x,y}) and 
$j_{\alpha}$ is a 
current operator along the $\alpha$-direction defined as 
$
j_{\alpha} = - {\rm i }\sum_{i,\sigma}  t
(
 c^{\dagger}_{i\sigma}             c_{i+\delta_{\alpha} ,\sigma}
- c^{\dagger}_{i+\delta_{\alpha} ,\sigma} c_{i\sigma}
),
$ 
where $\delta_{\alpha}$ is the unit vector along the $\alpha$-direction. 
$|n\rangle$ denotes an eigenstate with the energy eigenvalue of $E_n$. 
The ground state is represented by $n=0$. 
The averaging operation is performed to handle 
the anisotropic results due to the mixed BC and to reduce the FS correction.  
The Drude weight can be obtained from the combination of 
$\sigma (\omega)$ and the sum rule of 
$
\int_{0}^{\infty} \sigma (\omega) d\omega = 
\pi e^2 K ,
$
where 4$K$ denotes the kinetic energy per site. 
Hereafter we will call $K$ the total weight. 
When $U$ is large enough, 
we also calculate an effective carrier density defined as 
$
N_{\rm eff} = \frac{1}{\pi e^2} \int_{0}^{\omega_{\rm c}}
\sigma (\omega) d\omega ,
$
where $\omega_{\rm c}$ is a frequency just below 
the upper-Hubbard (UH) band. 
If $U$ is small, weight transfer to the 
UH band and the one within the lower-Hubbard band merge and 
$N_{\rm eff}$ is not well defined.  
To find $\omega_{\rm c}$ definitely, 
we take a large value of $U/t=16$.  
Note that, in the $t$-$J$ limit of {\it U}$\rightarrow\!\!\infty$, 
$N_{\rm eff}$ is equal to $K$. 
Besides, one can expect that the large value of $U$ would 
reduce the FS effect. 
In this work, we make a further procedure 
to reduce the FS effect. 
The procedure is to multiply a correction factor 
defined by $r=K_{\infty}/K_{N_{\rm s}}$, where 
$K_{\infty}$ and $K_{N_{\rm s}}$ are the total weights for $U$=0 
in the thermodynamic limit and of the $N_{\rm s}$-site system, 
respectively. 
Hereafter, all the quantities obtained after this procedure 
are labeled with suffix c, for example, $D^{\rm c}$. 
To check the validity of our calculations, we show results 
for the 1D Hubbard chain with $U/t=16$ 
in the inset of Fig.\mbox{\,}\ref{fig2}.  
It is seen that our Drude weight excellently agrees 
with the Bethe-ansatz result 
in the thermodynamic limit\cite{Kawakami_Yang}. 
Here one can also see the coherent feature, namely,  
$N_{\rm eff}^{\rm c}-D^{\rm c} \ll N_{\rm eff}^{\rm c}$ 
in the whole region of $\delta$.



Now, we present the results for the 2D Hubbard model with $U/t=16$ 
in Fig. \ref{fig2}. 
At half filling, the Drude weight is positive and finite 
but very small. 
This small and finite value comes from the FS effect and 
is expected to converge to zero 
if systems become larger. 
Actually, we have obtained $D^{\rm c}$\mbox{}$\sim$0.028 
for $\sqrt{10}\times$\mbox{}$\sqrt{10}$ sites for the same $U$ 
at half filling whereas $D^{\rm c}$\mbox{}$\sim$0.008
for $4\times 4$ sites, which shows a rapid convergence to zero.  
In the dilute-electron-density region ($\delta$\mbox{}$\sim$1), 
the interaction effect is small and 
the three quantities are close to the {\it U}=0 case. 
At larger electron density, 
the interaction works to prevent the electron motion 
and the three quantities decrease.  
In the whole region of $\delta$, 
$N_{\rm eff}$ exhibits a smooth and convex curve.  
In contrast with the results in Ref.\mbox{\,}\ref{Dagotto_Hub},  
compared to $N_{\rm eff}$, 
$D$ shows a concave shape 
below $\delta=0.25$ 
as shown in the data point closest to the Mott transition 
while  an overall convex behavior 
for higher $\delta$.  
This makes a quite asymmetric curve of $D$ 
between $\delta$\mbox{\,}$<$\mbox{\,}0.5 and $\delta$\mbox{\,}$>$\mbox{\,}0.5 
under reflection with respect to $\delta$=0.5. 
The concave shape for small $\delta$ suggests $z>2$.  
Although the number of data point at small $\delta$ is not sufficient 
to draw a definite conclusion on the exponent, the present result 
obtained after carefully reducing the FS effect 
supports that, as $\delta \rightarrow 0$, $D$ vanishes faster than 
linear $\delta$-dependence, and does not contradict $D\propto \delta^2$.   
The exponent deduced from 
$D\propto \delta^2$ is consistent with the ST 
characterized by $z=4$, and agrees with other results 
from the QMC method\cite{Furukawa_Imada,Assaad_Imada}.


Let us discuss the frequency dependence of the optical conductivity 
for $U/t$=16 
shown in Fig.\mbox{\,}\ref{fig3}(a)-(d). 
At half filling, only the excitations to the UH band 
are seen in Fig.\mbox{\,}\ref{fig3}(a). 
Upon increasing $\delta$, weights are transferred 
to the region below the Hubbard gap. 
In Fig.\mbox{\,}\ref{fig3}(b), one can see the large incoherent 
part below the UH band. 
The mid-gap incoherence has a larger weight on the lower-frequency side. 
As the frequency is increased, 
the height of the incoherence gradually decreases 
after the highest peak at about $\omega / t$\mbox{\,}$\sim$\mbox{\,}2. 
It is tempting to relate this structure to the shape of the $1/\omega$ tail 
seen in the experiment of high-$T_{\rm c}$ compound, 
although the incoherent response in the small frequency side 
$\omega /t$\mbox{\,}$<$2 clearly suffers from the FS effect. 
A similar behavior is seen in the study of the 2D $t$-$J$ 
model\cite{Poilblanc,Dagotto_review,Jaklic_Prelovsek}.  
Note that the data which Dagotto {\it et al.} reported 
in Ref.\mbox{\,}\ref{Dagotto_Hub} 
did not show such structure because the mid-gap incoherence itself 
is too small at the same $\delta$=0.125.  
Note also that the present calculations show that 
the incoherent peak gradually loses its weight 
with increasing $\delta$ 
while the long-tail structure survives. 


In summary, we have investigated the charge dynamics 
of the 2D Hubbard model. 
Lancz$\ddot{\rm o}$s diagonalization results under the careful 
treatment of the boundary condition have been presented. 
Near the metal-insulator transition, the optical conductivity has 
the large incoherence in the mid-gap region. 
At that doping concentration, 
the Drude weight is clearly suppressed, which 
supports the $z=4$ universality class of the Mott transition.

The authors thank H. Tsunetsugu and  Y. Motome 
for fruitful discussions. 
This work is supported by `Research for the Future Program' from Japan 
Society for Promotion of Science (JSPS-RFTF 97P01103). 
Computations was partly performed using the 
facilities of the Supercomputer Center, 
Institute for Solid State Physics (ISSP), University of Tokyo.  
Parallel calculations were done 
at FUJITSU VPP500 in ISSP and VPP700 in Kyushu University.

\vspace{-1cm}
\begin{figure}[h]
\caption{Drude weight, effective carrier density and total weight 
of the 2D Hubbard model when $U/t=16$. 
Inset shows the 1D results 
for 10 and 12 sites for $U/t=16$ 
together with the Bethe-ansatz result 
in the thermodynamic limit\cite{Kawakami_Yang} (solid curve) 
for comparison. 
}
\label{fig2}
\end{figure}

\vspace{-2cm}
\begin{figure}[h]
\caption{Incoherent part 
of $\sigma (\omega)$ for (a) $\delta$=0 (anti-periodic BC), 
(b) $\delta$=0.125 (mixed), (c) $\delta$=0.25 (mixed), 
(d) $\delta$=0.375 (periodic). 
Delta functions in the peaks are broadened with width of $\epsilon$=0.05. 
}
\label{fig3}
\end{figure}


\begin{thebibliography}{99}
\bibitem{mixed}A mixed BC in 2D is composed 
of the anti-periodicity along $x$-direction and the periodicity 
along $y$-direction, or the opposite combination. 
It corresponds to the case when 
the Aharonov-Bohm flux $\Phi$ in the hopping term 
introduced as the twisted BC  
is ($\pi$,0) or ($0,\pi)$. 
In Ref. \ref{Poilblanc}, it was reported that there exist cases where  
$\Phi$=($\pi$,0) or ($0,\pi)$ gives the lowest energy 
in the 2D $t$-$J$ model.
\end{thebibliography}
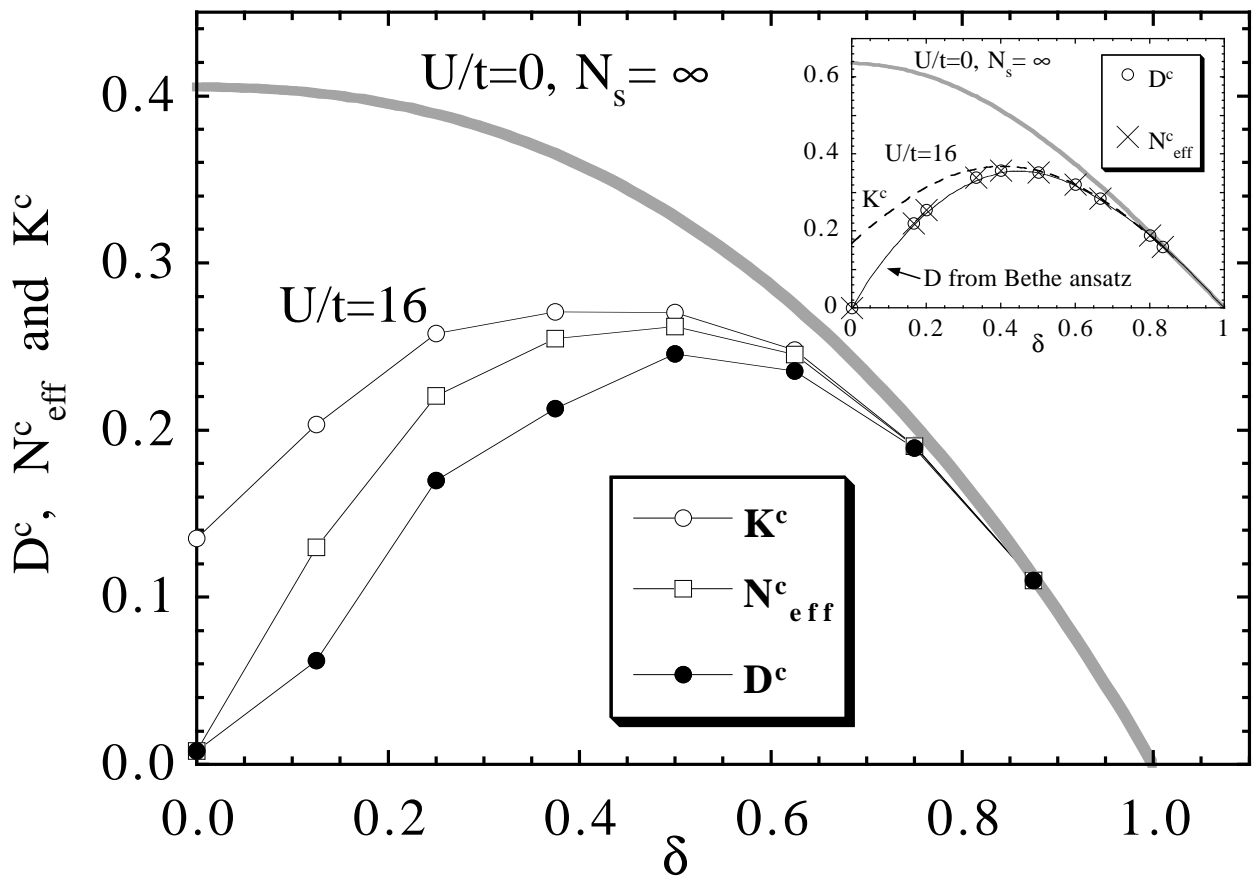
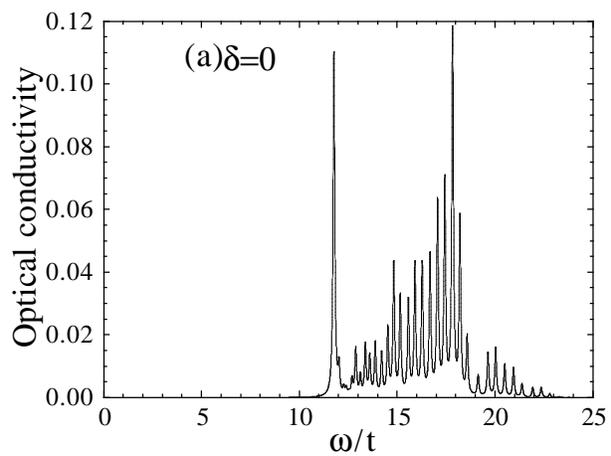
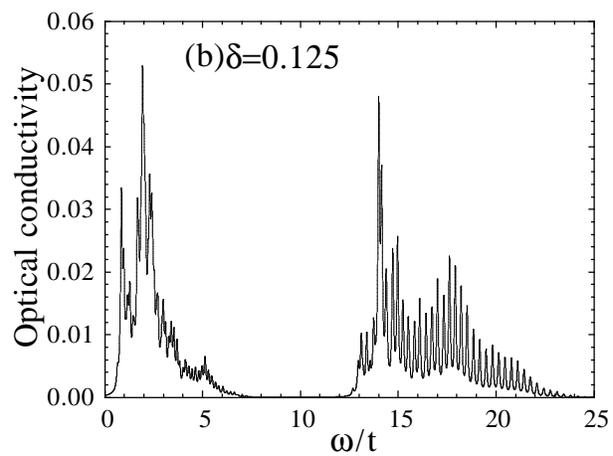
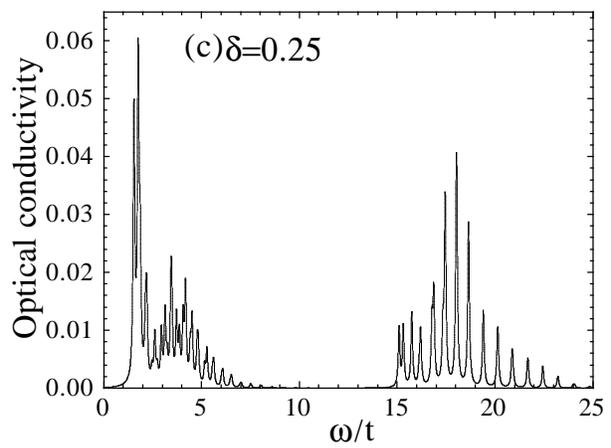
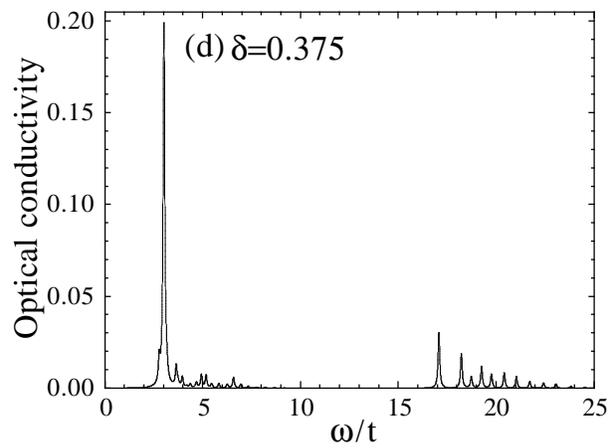
\end{document}